\documentclass{article}

\usepackage{arxiv}

\usepackage[utf8]{inputenc} % allow utf-8 input
\usepackage[T1]{fontenc}    % use 8-bit T1 fonts
\usepackage{hyperref}       % hyperlinks
\usepackage{url}            % simple URL typesetting
\usepackage{booktabs}       % professional-quality tables
\usepackage{amsfonts}       % blackboard math symbols
\usepackage{nicefrac}       % compact symbols for 1/2, etc.
\usepackage{microtype}      % microtypography
\usepackage{lipsum}		% Can be removed after putting your text content
\usepackage{graphicx}
\usepackage{natbib}
\usepackage{doi}
\usepackage{bm}
\usepackage[varvw]{newtxmath}       % selects Times Roman as basic font
\usepackage{amsmath}
\mathchardef\mhyphen="2D % Define a "math hyphen"
\usepackage{booktabs}
\usepackage{multirow}
\usepackage{comment}
\usepackage{adjustbox}
\usepackage{lscape}
\usepackage{longtable}
\usepackage{booktabs}
\usepackage{multirow}
\usepackage{threeparttable}
\usepackage[table,xcdraw]{xcolor}
\usepackage[
singlelinecheck=false % <-- important
]{caption}

\title{AI-driven non-intrusive uncertainty quantification of advanced nuclear fuels for digital twin-enabling technology}

\author{ {Kazuma ~Kobayashi}\\ 
	Nuclear, Plasma \& Radiological Engineering\\
	University of Illinois at Urbana-Champaign\\
	Urnaba, IL 61801, USA \\
        \And
	{Dinesh ~Kumar } \\
	Department of Mechanical Engineering\\
	University of Bristol\\
	Bristol BS8 1TR, UK \\
 \And
      {Syed Bahauddin ~Alam} \\
	Nuclear, Plasma \& Radiological Engineering\\
	University of Illinois at Urbana-Champaign\\
	Urnaba, IL 61801, USA \\
 }

\renewcommand{\shorttitle}{\textit{UQ/SA for Digital Twin} }

\hypersetup{
pdftitle={A template for the arxiv style},
pdfsubject={q-bio.NC, q-bio.QM},
pdfauthor={David S.~Hippocampus, Elias D.~Striatum},
pdfkeywords={First keyword, Second keyword, More},
}

\begin{document}
\maketitle

\begin{abstract}
In response to the urgent need to establish Digital Twin (DT) technology within next-generation nuclear systems, advancements in modeling methods and simulation codes are necessary. The increased complexity of models demands significant computational resources to quantify their uncertainties. To address this challenge, a non-intrusive uncertainty quantification method via polynomial chaos expansion is introduced as an efficient strategy within the finite element analysis-based fuel performance code BISON. Models of $\rm UO_2$ and $\rm U_{3}Si_{2}$ fuels, alongside SiC/SiC cladding material, were prepared to demonstrate the proposed method. The impact of four independent uncertain input variables on the system output was quantified, requiring fewer than 100 BISON simulations for each model. This approach not only accelerates the modeling and simulation task but also enhances the reliability in the development of DT-enabling technologies.
\end{abstract}

% section
\section{Introduction}
Since the Fukushima nuclear accident in 2011, the environment surrounding the nuclear sector has changed significantly. There is a renewed focus on extending the service life of existing reactors, optimizing their operation, and increasing their efficiency. Accordingly, developing accident-tolerant fuels (ATFs) is one of the prime topics \cite{almutairi2022weight}. Against these trends, the U.S. Nuclear Regulatory Commission (NRC) advocated the application of digital twin (DT) technology to the nuclear field as an area of future research \cite{yadav2021technical,yadav2022aug,yadav2023feb,yadav2023sep,yadav2023technical}. 

Per the NRC \cite{yadav2021technical}, Digital Transformation in nuclear power plants is anticipated to yield several advantages, including improved operational efficiency, heightened safety and reliability, fewer errors, quicker dissemination of information, and superior forecasting capabilities. However, DT is an entirely new concept in the nuclear field, and the current stage of development is to identify the goals to be achieved and the technical issues to be addressed. 

DT can be categorized by purpose, but the components are the same: visualization, data processing, system update, prediction, and decision-making  \cite{yu2022hybrid,chen2022digital,bondarenko2020development}. In this context, ``visualization" is used for preparing the virtual asset of a physical system and visualizing it on the computer. It is the foundation for building DT. It is similar to conventional simulation; imagine commercial software such as computer-aided design (CAD). Second, ``data processing" is responsible for transferring the physical system's sensor data to the digital assets on the computer. Large systems such as nuclear reactors are expected to have a large number of sensors and data size and, therefore, require constructing an appropriate database. The third ``system update" clearly differs between DT and traditional simulation. A typical simulation predicts the system state at a given time, assuming the system parameters are known. However, the objective of DT is to monitor and predict system conditions over a long time scale from the start of operation of that system to its shutdown (e.g., from months to years). It means that the system parameters must be treated as a function of their system operation time, and their values must be updated as time evolves. Obtaining the system parameters at a given time is generally classified as an \textit{inverse} problem, and the ``update" handles the sequence of updating the obtained values at the next time step. The ``prediction" predicts the system state using the above-mentioned updated system parameters. This process is a \textit{forward} problem to solve for the system state, and the user can select any solvers according to the information they want to obtain \cite{kobayashi2024deep, kobayashi2024improved}. For example, for industrial products such as automobiles and aircraft, commercial FEA tools such as ABAQUS and ANSYS or in-house codes owned by each vendor would be an option. The final ``decision-making" makes decisions about system maintenance, modifications, requests for maintenance, etc., based on the results of the previous forecasting module \cite{kobayashi2023surrogate, kobayashi2022practical}. It is challenging since it involves not only the design values of the system but also the restrictions imposed by national and international conventions. For example, consider an automobile as a system. The vehicle must meet each country's exhaust gas emission regulations even if the driving performance is at the expected value. Therefore, even if the same system is adopted, changing the decision-making module becomes a point of caution. All of these modules are essential, but in a nuclear system, the prediction tools are independent for each of their applications. In this study, the nuclear fuel performance evaluation code BISON is assumed to be one of the prediction tools in nuclear DT, and its potential applications are explored.

BISON, a nuclear fuel performance code developed by the Idaho National Laboratory (INL), utilizes finite element analysis to accommodate a variety of fuel types, including metallic rod (U-Pu-Zr) and plate fuel (U-10Mo), TRISO particle fuel, and fuel rods for light water reactors ($\rm UO_{2}$) \cite{hales2016bison, bison_general}. Constructed on the Multiphysics Object-Oriented Simulation Environment (MOOSE) framework \cite{permann2020moose}, BISON supports analyses for both 2D axisymmetric and 3D geometries. It is adept at solving fully coupled thermomechanical and species diffusion equations. BISON is capable of modeling temperature and burnup-dependent thermal properties, as well as fission product swelling, densification, thermal and irradiation creep, fracture, and fission gas production/release, showcasing the interplay of these phenomena through tensor mechanics \cite{hales2016bison}. Plasticity, radiation growth, fuel cladding chemical interaction (FCCI), damage mechanics, and thermal and radiation creep models are used in addition to general material properties for clad materials. Some models represent mechanical contact, gap heat transfer, change in gap/plenum pressure with plenum volume, gas temperature, and accumulation of fission gas \cite{hales2016bison}. BISON is used to evaluate the condition of nuclear fuel pellets and cladding during reactor operation and is expected to be adopted as a predictive tool in nuclear system DT. However, the prediction is based on system parameters inferred from sensor data to obtain the system state. No sensor (for temporal synchronization within DT) exists with perfect accuracy, and uncertainties exist in its data. Therefore, quantifying the impact of uncertain input values on system response is a prerequisite for prediction tools \cite{verma2023reliability, kobayashi2023data,amadeh2022quantifying,zhang2020quantification,zhou2020uncertainty}. 

From a DT perspective, the development of novel Accident-Tolerant Fuel (ATF) technology encounters several challenges, including (i) data unavailability, (ii) lack of data, missing data, and inconsistencies in data, and (iii) model uncertainty. Overcoming these hurdles is critical to establishing trust in the development of DT frameworks. It's essential to note that DT-enabling technologies encompass three main domains as identified by \cite{yadav2021technical}: (i) Modeling \& Simulation (M\&S), which includes uncertainty quantification (UQ) and data analytics \cite{kumar2019influence, kumar2022multi} through trustworthy Artificial Intelligence/Machine Learning (AI/ML) \cite{kobayashi2024improved,kobayashi2024deep,kobayashi2024explainable}, physics-based models, and data-informed modeling; (ii) advanced sensors/instrumentation with real-time signal processing~\citep{kabir2010non,kabir2010theory,kabir2010watermarking,kabir2010loss} ; and (iii) data \& information management. Among these, UQ is a crucial aspect of DT-enabling technologies that is vital for ensuring trustworthiness, a component that BISON is required to integrate to fulfill DT prerequisites. This study explores the integration of polynomial chaos expansion (PCE)-based UQ employing a non-intrusive approach within BISON to address the M\&S requirements of the Accelerated Fuel Qualification (AFQ) for ATF. The innovation of this research lies in its application of a non-intrusive, computationally efficient polynomial chaos-based UQ methodology within the BISON code for sophisticated ATF concepts. Moreover, this research marks the inaugural effort to implement and analyze uncertainty estimations for DT-enabling technology.

It should be noted that a book chapter \cite{kobayashi2023uncertainty} has been exclusively dedicated to the UO$_{2}$+SiC/SiC system case study (A preprint is also available online). Conversely, this article provides a comprehensive elaboration on the methodologies, the entire process framework for the UQ (Uncertainty Quantification/Sensitivity Analysis) approach, and the relevant outcomes for both U$_{3}$Si$_{2}$+SiC/SiC and UO$_{2}$+SiC/SiC systems, in the context of facilitating digital twin technology.

\section{Application of polynomial chaos expansion for uncertainty quantification}
The technique of Polynomial Chaos Expansion (PCE) is an established method for uncertainty quantification (UQ) that has proven effective for stochastic simulations, as highlighted in various studies \cite{daroczy2016analysis,tang2020uncertainty}. PCE allows for the characterization of variables and the solution output through mean, variance, higher-order moments, and probability density functions \cite{Kumar2016, Kumar2020a}. It is important to note that our previous work \cite{kobayashi2023uncertainty} has already laid out the foundational methodology for constructing multi-dimensional polynomials (Section \ref{sec:mult_pols}), as well as the regression technique for estimating Polynomial Chaos coefficients (Section \ref{sec:regression}).

% explanation of orthogonality and jacobi & legendre
Orthogonal polynomials constitute a class of polynomials demonstrating orthogonality with respect to a specific weight function. For applications in the stochastic domain on computational platforms, Hermite, Laguerre, Jacobi, and Legendre polynomials find frequent use. Orthogonality, in this context, is formally defined as: 

$$\int_{\xi}\psi_{i}(\xi)\psi_{j}(\xi)W_{\xi}(\xi)d\xi = \left< \psi_{i}\psi_{j} \right> = \delta_{ij} \left< \psi_{i}^{2} \right>$$

\begin{figure}[!htbp]
    \centering
    \includegraphics[width=14cm]{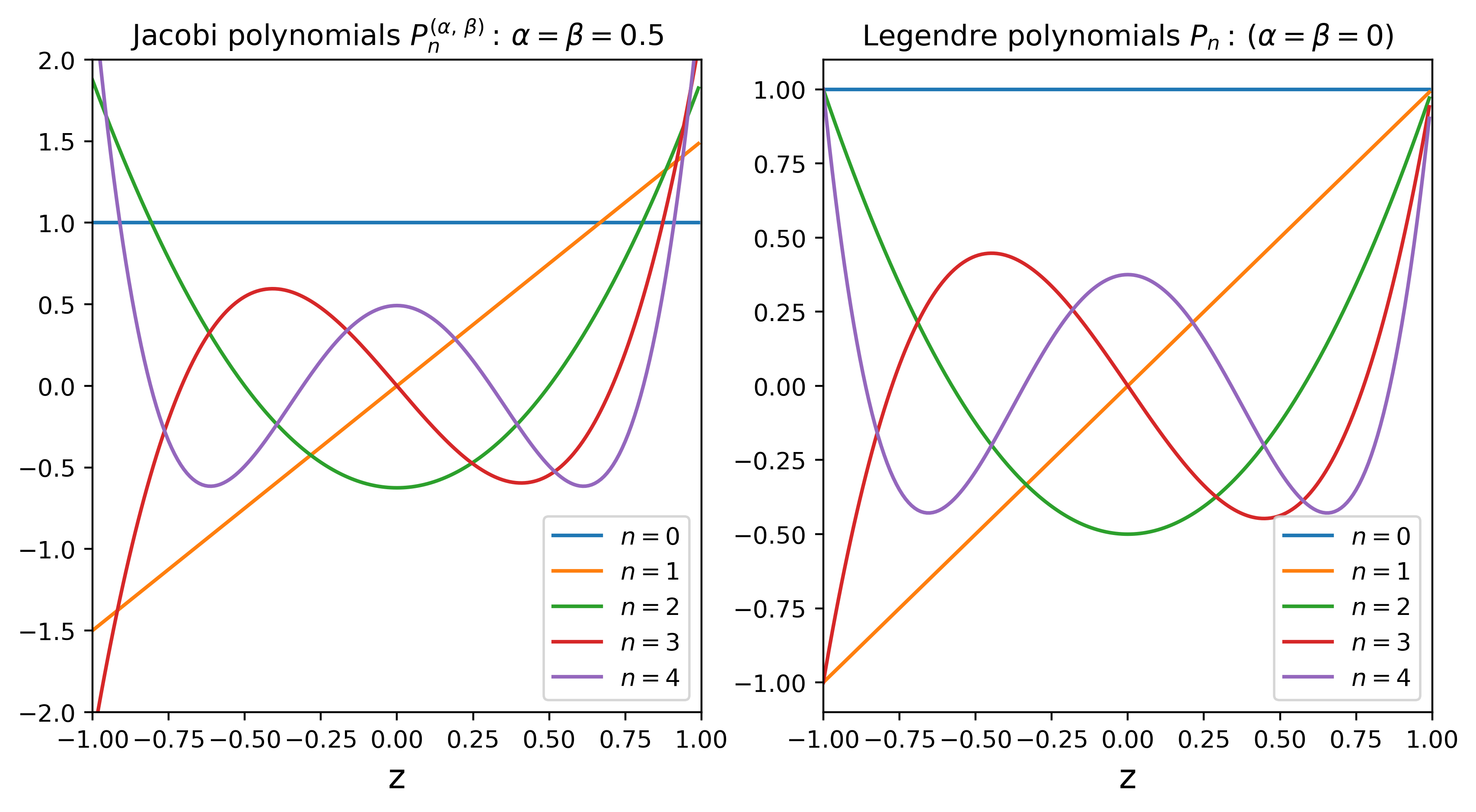}
    \caption{Example of 1D Jacobi polynomials and Legendre polynomials. $n$ represents the order of polynomials.}
    \label{fig:jacobi_legendre}
\end{figure}

In this definition, $W_{\xi}(\xi)$ signifies the probability distribution function (or alternatively, the weighting function) associated with the random variable $\xi$, $\delta_{ij}$ denotes the Kronecker delta, $\psi_{i}(\xi)$ represent the polynomial basis functions, and $\left< \psi_{i}\psi_{j} \right>$ stands for the inner product. This formal presentation of orthogonality underscores the fundamental property that when the product of distinct orthogonal polynomials is integrated over a predefined interval (with the weight function incorporated), the result is zero. This characteristic makes orthogonal polynomials exceptionally valuable for a range of numerical and analytical tasks within probability theory, computational simulations, and differential equation modeling. Jacobi polynomials are implemented as default basis functions. The Jacobi polynomials are defined by Rodrigues' formula:

\begin{equation}
    P_{n}^{(\alpha,\beta)}(z) = \frac{(-1)^{n}}{2^{n}n!}(1-z)^{-\alpha}(1+z)^{-\beta}\frac{d^{n}}{dz^{n}}\bigl\{(1-z)^{\alpha}(1+z)^{\beta}(1-z^{2})^{n}\bigl\}
\end{equation}
with the constraints of $\alpha,\beta > -1$. Also, the orthogonality can be produced by:

\begin{equation}
\int_{-1}^{1} P_{n}^{(\alpha,\beta)}(z) P_{m}^{(\alpha,\beta)}(z)(1-z)^{\alpha}(1+z)^{\beta}\,dz = \frac{2^{\alpha+\beta+1}}{2n+\alpha+\beta+1}\frac{\Gamma(n+\alpha+1)\Gamma(n+\beta+1)}{\Gamma(n+\alpha+\beta+1)}\delta_{n,m}
\end{equation}
with the beta density $w(z)=(1-z)^{\alpha}(1+z)^{\beta}$ on $[-1,1]$. By setting the parameters $\alpha=\beta=0$ in this study, the Jacobi polynomials are modified to Legendre polynomials as follows:
\begin{equation}
    P_{n}(z) = \frac{1}{2^{n}n!}\frac{d^{n}}{dz^{n}}(z^{2}-1)^{n}
\end{equation}
The shapes of the Jacobi and Legendre polynomials are illustrated in Figure \ref{fig:jacobi_legendre}.

%%%%%
In polynomial chaos methods, it's feasible to distinguish between deterministic and non-deterministic orthogonal polynomials. Consider the breakdown of a stochastic variable $u(x, \xi)$ as follows:
\begin{equation}
\label{eq:decompose_rephrased}
    u(x, \xi) = \sum_{i=0}^{P} u_{i}(x)\psi_{i}(\xi)
\end{equation}
Here, $u_{i}(x)$ represents the deterministic coefficients of expansion, and the series includes $P+1$ terms in total. The expected value of $u(x)$ is represented as:
\begin{equation}
\label{eq:mean_rephrased}
    E[u] = u_{0}
\end{equation}
Furthermore, the variance can be expressed as:
\begin{equation}
\label{eq:var_rephrased}
    E[(u- E[u])^2] = \sigma_{u}^{2} = \sum_{i=1}^{P}u_{i}^{2} \left< \psi_{i}^{2} \right>
\end{equation}
This delineation allows for a clear separation between the deterministic and stochastic components within the polynomial chaos framework.

Simulations often involve 20 or more unknown parameters. When applying PCE, each original parameter is replaced by a set of additional parameters. For example, 6 original parameters might expand into 120 parameters after PCE. This expansion makes computations significantly more demanding, especially as the initial number of uncertain parameters increases. 

The rapid escalation in computational demand with increasing dimensions, termed the "curse of dimensionality," stands as the foremost limitation of polynomial chaos methods. This challenge propels researchers to seek out and innovate more streamlined stochastic models tailored for uncertainty quantification within intricate industrial settings \cite{Kumar2016}.

\subsection{Multi-dimensional polynomials}
\label{sec:mult_pols}
Constructing Multi-dimensional polynomials from their one-dimensional counterparts is crucial for examining the effects of stochastic variables on outcomes. The PCE of a multi-dimensional order $p$ can be delineated using 1D polynomials. For illustration, a 2D PCE with an order of $p=3$ is selected. The methodology for constructing multi-dimensional polynomials was also detailed in our earlier research \cite{kobayashi2023uncertainty}.

When we define a set of 1-dimensional orthogonal polynomials for a order of 3 as $\{ \Psi_{0},\Psi_{1}, \Psi_{2}, \Psi_{3} \}$, a 2D stochastic variable $u(\xi_{1}, \xi_{2})$ can be represented as a sum of these polynomial functions and their respective coefficients.

\begin{equation}
    \begin{split}
        u(\xi_{1},\xi_{2}) &= u_{00}\Psi_{0}\\
        &+ u_{10}\Psi_{1}(\xi_{1}) + u_{01}\Psi_{1}(\xi_{2})\\
        &+ u_{20}\Psi_{2}(\xi_{1}) + u_{11}\Psi_{1}(\xi_{1})\Psi_{2}(\xi_{2}) + u_{02}\Psi_{2}(\xi_{2})\\
        &+ u_{30}\Psi_{3}(\xi_{1}) + u_{21}\Psi_{2}(\xi_{1})\Psi_{1}(\xi_{2}) + u_{12}\Psi_{1}(\xi_{1})\Psi_{2}(\xi_{2}) + u_{03}\Psi_{3}(\xi_{2}).
    \end{split}
\end{equation}

For a collection of multi-dimensional independent variables, $\boldsymbol{\xi} = (\xi_{1},...,\xi_{n})$, the probability density function (PDF) is defined as follows:

\begin{equation}
\label{eq:weight_rephrased}
    \boldsymbol{W}(\boldsymbol{\xi}) = \prod_{i=1}^{n}W_{i}(\xi_i)
\end{equation}
Here, $W_{i}(\xi_i)$ represents the individual PDF for each random variable $\xi_{i}$ \cite{Kumar2016, Kumar2020a}. The total number of polynomial terms $P+1$ in Eq. \ref{eq:decompose_rephrased} is related to the polynomial order ($p$) and the count of input variables ($n$) as:

\begin{equation}
\label{eq:tot_pols_rephrased}
    P+1 = \frac{(p+n)!}{p!n!}
\end{equation}
Consequently, for a polynomial order of $p=3$ and $n=2$ variables, the total number of polynomial terms ($P+1=10$) is determined.

%% PCE

\subsection{Estimation of coefficients}
\label{sec:regression}
One common approach for determining the unknown set of polynomial coefficients in polynomial chaos expansions is regression analysis. This statistical method estimates the relationship between a dependent variable and one or more independent variables. Our previous study delves into the fundamental regression approach for estimating coefficients \cite{kobayashi2023uncertainty}. The most typical type, linear regression, finds the line that best fits the data. Linear regression is widely used in data analysis and surrogate modeling methods, including Gaussian processes \cite{kobayashi2022practical}.

Walters' regression-based non-intrusive polynomial chaos method \cite{Kumar2020a, Walters2003}  provides a computationally efficient way to calculate the polynomial coefficients needed for PCEs. This practical approach to uncertainty quantification works as follows:

\begin{equation}
    u(x;\boldsymbol{\xi}) = \sum_{i=0}^{P}u_{i}(x)\psi_{i}(\boldsymbol{\xi})
\end{equation}
where $u(x;\boldsymbol{\xi})$ represents stochastic quantity of interest.
Using $m$ samples $(\boldsymbol{\xi^{j}}= \{ \xi_{1},...,\xi_{n_{s}} \}^{j}; j=1,...,m)$ shown by the PDF $\mathbf{W}(\boldsymbol{\xi})$ from Eq. \ref{eq:weight_rephrased} and the corresponding model output  $u(x;\boldsymbol{\xi}^{j})$, this system can be represented as a matrix equation:

\begin{equation}
\begin{bmatrix}
\psi_{0}(\boldsymbol{\xi}^{1}) && \psi_{1}(\boldsymbol{\xi}^{1}) && \cdots && \psi_{P}(\boldsymbol{\xi}^{1}) \\
\psi_{0}(\boldsymbol{\xi}^{2}) && \psi_{1}(\boldsymbol{\xi}^{2}) && \cdots && \psi_{P}(\boldsymbol{\xi}^{2}) \\
\vdots  && \vdots && && \vdots\\
\psi_{0}(\boldsymbol{\xi}^{m}) && \psi_{1}(\boldsymbol{\xi}^{m}) && \cdots && \psi_{P}(\boldsymbol{\xi}^{m}) \\
\end{bmatrix}
\begin{bmatrix}
 {u_{0}}(x) \\ {u_{1}}(x) \\ \vdots \\ {u_{P}}(x)
\end{bmatrix}
=
    \begin{bmatrix}
u(x;\boldsymbol{\xi}^{1}) \\ u(x;\boldsymbol{\xi}^{2}) \\ \vdots \\ u(x;\boldsymbol{\xi}^{m})
    \end{bmatrix}
\end{equation}
or in matrix notation

\begin{equation}
\label{eq:system}
    [A]\{u\} = \{b\}
\end{equation}
Our primary goal is to solve the equation for $\{u\}$. Assuming $m > P + 1$, we can express this in matrix notation:

\begin{equation}
\label{eq:f_matrix}
  \{u\} = \left( [A]^{\top}[A] \right)^{-1}[A]^{\top}\{b\}
\end{equation}

With this solution, we can leverage Eqs. \ref{eq:mean_rephrased}, \ref{eq:var_rephrased}, and \ref{eq:weight_rephrased} to quantitatively assess the mean and variance of system outputs. This integrated approach allows us to directly account for the impact of uncertainty in our input variables. 
% Sampling
In the preceding system of equations, the matrix $[A]$ is referred to as the design matrix because it contains information about the polynomial values that the design samples \cite{Kumar2016}. In least-squares-based regression, the design matrix $[A]$ plays a pivotal role. Therefore, the sampling method and the number of samples significantly affect the instancing of the design matrix. Various sampling strategies, such as the Latin hypercube \cite{helton2003latin}, the Sobol sequence \cite{sobol1967distribution}, and random sampling \cite{etikan2017sampling} can be used to build the design matrix. The influence of various sampling techniques and the number of total samples was studied in the previous works \cite{Kumar2020,hosder2006non, kumar2021quantitative,kumar2022multi}, and it was concluded that more than twice as many samples, $2(P+1)$, as coefficients are needed to ensure accuracy.

\section{Methods}
\label{sec:models}
This section outlines the material models and setup required for computational tasks in BISON, focusing on demonstrating uncertainty quantification analysis. The materials selected for these sample problems include $\rm UO_{2}$/$\rm U_{3}Si_{2}$ for fuels and SiC/SiC for cladding. It's important to note that this study aims to apply the PCE technique to BISON, and to streamline the analysis, the mass density and thermal conductivity values for both $\rm UO_{2}$/$\rm U_{3}Si_{2}$ fuels and SiC/SiC cladding are held constant.

Despite the focus on PCE application, the inherent thermal conductivity models for $\rm UO_{2}$/$\rm U_{3}Si_{2}$ and SiC/SiC within BISON are not utilized in this study. Nevertheless, a brief introduction to these available models is provided.

\subsection{UO$_{2}$ Models}
Material models are available in BISON for the following material properties of $\rm UO_{2}$: thermal conductivity, specific heat, elasticity tensor, creep, thermal expansion, and swelling.

The thermal conductivity and specific heat capacity of $\rm UO_{2}$ can be determined using various empirical models such as those proposed by Fink-Lucuta \cite{fink00, Lucuta1996}, Halden \cite{lanning2005}, NFIR \cite{nrc_nfir_ltr, anatech_NFIR_Gd_ltr}, Modified NFI \cite{ohira97}, Ronchi-Staicu \cite{ronchi2004, staicu2014}. These models account for factors like temperature, porosity, burnup, stoichiometry, and the presence of gadolinium, affecting the thermal properties of $\rm UO_{2}$.

While the thermal conductivity models provide a range of semi-empirical and analytical methods, the specific heat capacity is described only through empirical models. The formula for specific heat, $C_{p}$ (J/kg-K), is given by

\begin{equation}
\label{eq:c_p_uo2}
    C_{p} = \frac{K_{1}\theta^{2}\text{exp}(\frac{\theta}{T})}{T^{2}[\text{exp}(\frac{\theta}{T})-1]^{2}} + K_{2}T + \frac{YK_{3}E_{D}}{2RT^{2}}\text{exp}\left( -\frac{E_{D}}{RT} \right)
\end{equation}
Here, $T$ denotes the temperature in Kelvin, $Y$ the oxygen-to-metal ratio, $R$ the universal gas constant (8.3145 J/mol-K), and the coefficients $K_{1}$, $K_{2}$, $K_{3}$, $\theta$, and $E_{D}$ are empirically determined parameters. The standard values for these parameters are outlined in Table \ref{tab:cp_uo2}. In this research, while the thermal conductivity is chosen based on user-defined criteria, the specific heat employs the empirical model delineated in Eq. \ref{eq:c_p_uo2}.

\begin{table}[htbp]
\centering
\caption{Constant parameters for $\rm UO_{2}$ specific heat retrieved from \cite{luscher2015}}
\label{tab:cp_uo2}
\begin{tabular}{@{}lccc@{}}
\toprule
Constant & Value  & Units \\ \midrule
$K_1$       & 296.7  & $\rm J/kg \mhyphen K$      \\
$K_2$       & $2.43 \times 10^{-2}$  & $\rm J/kg \mhyphen K^{2}$  \\
$K_3$       & $8.745 \times 10^{7}$  & $\rm J/kg$        \\
$\theta$    & 535.285               & $\rm K$           \\
$Y$         & 2.0                   & -                 \\
$E_D$       & $1.577 \times 10^{5}$ & $\rm J/mol$       \\ \bottomrule
\end{tabular}
\end{table}

For the elasticity tensor model, the predefined constant values of Young's modulus and Poisson's ratio are set at $E = 2.0 \times 10^{11}$ Pa and $\nu = 0.345$ \cite{bison}, respectively. There is no change in these values across the material's entire temperature range of use with this setting.

The creep behavior of $\rm UO_{2}$ is characterized by the MATPRO FCREEP model \cite{hagrman93}, which accounts for both thermal creep and irradiation creep. The model is formulated as:

\begin{equation}
\label{eq:creep_uo2}
\dot{\epsilon} = \frac{A_{1} + A_{2} \dot{F}}{(A_{3}+D)G^{2}}\sigma \text{exp} \left( -\frac{Q_{1}}{RT} \right) + \frac{A_{4}}{(A_{6} + D)} \sigma^{4.5} \text{exp} \left( -\frac{Q_{2}}{RT} \right) + A_{7} \dot{F} \sigma
\end{equation}

In this equation, $\dot{\epsilon}$ represents the creep rate (1/s), $\sigma$ the effective stress (Pa), $T$ the temperature (K), $D$ the fuel density (expressed as a percentage of theoretical density), $G$ the grain size ($\rm \mu m$), $\dot{F}$ the volumetric fission rate ($\rm fissions/m^{3}$-s), $Q_{i}$ the activation energies (J/mol), and $R$ the universal gas constant (8.3143 J/mol-K). The coefficients used in Eq.\ref{eq:creep_uo2} are detailed in Table \ref{tab:creep_uo2_params}. Additionally, the model defines the activation energies $Q_{i}$ as functions of the oxygen-to-metal ratio $x$.

\begin{equation}
\begin{split}
    Q_{1} &= 74.829 f(x) + 301.762 \\
    Q_{2} &= 83.143 f(x) + 469.191
\end{split}
\end{equation}
where the function $f(x)$ is defined as
\begin{equation}
    f(x) = \frac{1}{\text{exp}\left( -\frac{20}{\ln(x-2)} -8 \right)+1}
\end{equation}

\begin{comment}
\begin{table}[htbp]
\centering
\caption{Constant parameters for the $\rm UO_{2}$ creep model \cite{hagrman93}}
\label{tab:creep_uo2_params}
\begin{adjustbox}{width=\textwidth}
\begin{tabular}{@{}lcccccc@{}}
\toprule
Parameter & $A_{1}$ & $A_{2}$ & $A_{3}$ & $A_{4}$ & $A_{6}$ & $A_{7}$   \\
Value     & 0.3919 & $1.3100 \times 10^{-19}$ & -87.7 & $2.0391\times 10^{-25}$ & -90.5 & $3.7226 \times 10^{-35}$\\ \bottomrule
\end{tabular}
\end{adjustbox}
\end{table}
\end{comment}

\begin{table}[htbp]
\centering
\caption{Constant parameters for the $\rm UO_{2}$ creep model \cite{hagrman93}}
\label{tab:creep_uo2_params}
\begin{tabular}{@{}lc@{}}
\toprule
Parameter & Value                    \\ \midrule
$A_{1}$  & 0.3919                   \\
$A_{2}$  & $1.3100 \times 10^{-19}$ \\
$A_{3}$  & -87.7                    \\
$A_{4}$  & $2.0391 \times 10^{-25}$ \\
$A_{6}$  & -90.5                    \\
$A_{7}$  & $3.7226 \times 10^{-35}$ \\ \bottomrule
\end{tabular}
\end{table}

%% Thermal expansion
For the thermal expansion of $\rm UO_{2}$, the MATPRO FTHEXP function was selected. This model can compute dimensional changes in unirradiated fuel pellets as a temperature-dependent function, and handles $\rm UO_{2}$ and $\rm PuO_{2}$ in solid, liquid, or both phases and includes solid-liquid expansion \cite{hagrman93}. The equation for the thermal expansion of $\rm UO_{2}$ in solid phase is expressed by

\begin{equation}
\label{eq:fthexp}
    \frac{\Delta L}{L_{0}} = P_{1}T - P_{2} + P_{3} \text{exp}\left( -\frac{P_{E_D}}{k_{B}T} \right)
\end{equation}
where $\Delta L/L_{0}$ is the linear strain caused by thermal expansion, $T$ is the temperature (K), $P_{E_D}$ is the energy of formation of a defect (J), $k_{B}$ is the Boltzmann's constant ($1.38\times 10^{-23}$ J/K). The constants $P_{E_D}$ and $P_{i}$ are determined by the fuel material. Values for $\rm UO_{2}$ and $\rm PuO_{2}$ are listed in Table \ref{tab:fthexp}. %The computed thermal strain is shown in Figure \ref{fig:fthexp}.

\begin{table}[htbp]
\centering
\caption{Constant parameters used in MATPRO FTHEXP function \cite{hagrman93}}
\label{tab:fthexp}
\begin{tabular}{@{}lccc@{}}
\toprule
Parameter & $\rm UO_{2}$ & $\rm PuO_{2}$ & Units \\ \midrule
$P_1$       & $1.0 \times 10^{-5}$  & $9.0 \times 10^{-6}$  & $\rm K^{-1}$ \\
$P_2$       & $3.0 \times 10^{-3}$  & $2.7 \times 10^{-3}$  & -      \\
$P_3$       & $4.0 \times 10^{-2}$  & $7.0 \times 10^{-2}$  & -      \\
$P_{E_D}$       & $6.9 \times 10^{-20}$ & $7.0 \times 10^{-20}$ & J      \\ \bottomrule
\end{tabular}
\end{table}

\begin{comment}
\begin{figure}[!htbp]
    \centering
    \includegraphics[width=10cm]{figures/FTHEXP.png}
    \caption{The thermal expansion strain of $\rm UO_{2}$ and $\rm PuO_{2}$ computed from MATPRO FTHEXP function}
    \label{fig:fthexp}
\end{figure}
\end{comment}

%% Volumetric Swelling
Fuel volume swelling is caused by densification and fission products. For densification, the ESCORE model \cite{rpt_rashid_2004} is selected. It is given by

\begin{equation}
\label{eq:escore}
    \left( \frac{\Delta V}{V_{0}}\right)_{den} = \Delta \rho_{0} \Biggr[ \text{exp}\left( \frac{Bu\,ln(0.01)}{C_{D}Bu_{D}}  \right)-1 \Biggr]
\end{equation}
where $\Delta V/V_{0}$ is the densification strain, $\Delta \rho_{0}$ is the total densification, $Bu$ is the burnup (fissions/atoms-U), and $Bu_{D}$ is the burnup at which densification is complete (fissions/atoms-U). $C_{D}$ represents the temperature dependence and changes value at 750 $^{\circ}C$. Hence, during the simulation, the parameter $C_{D}$ is automatically switched as follows:

\begin{equation}
    C_{D} = 
\begin{cases}
7.235-0.0086(T-25)/500 & (T < 750 \,  \rm{^{\circ}}C)\\
1.0 & (T \geq 750 \, \rm{^{\circ}}C)
\end{cases}
\end{equation}

\noindent
Fission products also contribute to fuel swelling, and they can be classified into two types: (1) solid fission products and (2) gaseous fission products. For both kinds of swellings, the MATPRO models are applicable. The model for solid fission product ($sfp$) is provided as a function of burnup:

\begin{equation}
    \Delta V_{sfp} = 5.577 \times 10^{-5} \rho \Delta Bu
\end{equation}
and for gaseous fission product ($gfp$):
\begin{equation}
\Delta V_{gfp} = 1.96 \times 10^{-31} \rho \Delta Bu (2800-T)^{11.73} \text{exp}\left( -0.0162(2800-T) \right) \text{exp}(-0.0178\rho Bu)
\end{equation}
where $\Delta V_{sfp}$ and $\Delta V_{gfp}$ represent the volumetric swelling increment, $\rho$ is the density ($\rm kg/m^{3}$) of fuel and $T$ is the temperature in Kelvin.

\subsection{U$_{3}$Si$_{2}$ Models}
The thermal conductivity and specific heat models for $\rm U_{3}Si_{2}$ are available in the BISON code. There are four options of thermal conductivity: WHITE \cite{white2015}, SHIMIZU \cite{Shimizu_1965}, ZHANG \cite{zhang2017thermal}, and HANDBOOK \cite{u3si2_handbook2}. These models compute thermal conductivity as a function of temperature, but only the Zhang model has silicon concentration as an additional variable. For the specific heat capacity of $\rm U_{3}Si_{2}$, three different models are available: WHITE \cite{white2015}, IAEA \cite{IAEA_tecdoc643}, and HANDBOOK \cite{u3si2_handbook2}. The heat capacity of these models is expressed as a linear function of temperature as follows:

\begin{equation}
    C_{p} = 
\begin{cases}
140.5 + 0.02582T & (\text{WHITE})\\
199 + 0.104(T-273.15) & (\text{IAEA}) \\
1000.0 (3.52 \times 10^{-5}T+0.18) & (\text{HANDBOOK})
\end{cases}
\end{equation}
As the same as the $\rm UO_{2}$ models, this study selects the user-defined thermal conductivity and sets the WHITE model to specific heat.

The elasticity tensor model is selected to compute Young's modulus and Poisson's ratio for $\rm U_{3}Si_{2}$. The model reported by White \cite{u3si2_handbook2} computes the values as a function of porosity. The Young's modulus $E$ (GPa) and shear modulus $G$ (GPa) are given by:

\begin{equation}
\label{eq:young_shear}
\begin{split}
    E &= -6.425p + 142.68 \\
    G &= -2.901p + 61.27
\end{split}
\end{equation}
where $p$ is the porosity (\%). Using these values, the Poisson's ratio $\nu$ can be computed from:
\begin{equation}
    \nu = \frac{E}{2.0G} - 1.0
\end{equation}
The porosity used in Eq. \ref{eq:young_shear} comes from the following:
\begin{equation}
    p = \left( 1.0 - \frac{\rho_{curt}}{\rho_{theor}} \right) \times 100\%
\end{equation}
where $\rho_{curt}$ is the current density and $\rho_{theor}$ is the theoretical density of 12200 $\rm kg/m^{3}$.

%% Thermal creep
In order to simulate the thermal creep of $\rm U_{3}Si_{2}$, Freeman \cite{freeman2018}, Metzger \cite{metzger2016}, and Yingling \cite{yingling2019} models are available in the BISON. The Yingling model is employed in this study to account for particle size. Using the model, the creep rate is given by:
\begin{equation}
\dot{\epsilon} = 4.841 \times 10^{-19} \sigma^{1.936} d^{-1.86} \text{exp} \left( \frac{223100}{RT} \right) 
\end{equation}
where $R$ is the ideal gas constant (8.3143 J/mol-K), $\sigma$ is the effective stress (Pa), $T$ is the temperature (K), and $d$ is the average grain size (m). For the value of $d$, the BISON default value of $2\times 10^{-5}$ (m) is selected.

%% Volume swelling (using ESCORE models) different is C_D
%\cite{rpt_rashid_2004}
Volumetric swelling due to densification and fission products occurs in $\rm U_{3}Si_{2}$. Assuming the $\rm U_{3}Si_{2}$ densification phenomena similar to $\rm UO_{2}$,  the ESCORE model (Eq. \ref{eq:escore}) is selected. The difference between $\rm U_{3}Si_{2}$ and $\rm UO_{2}$ in this model is the constant parameter $C_{D}$ as below:

\begin{equation}
    C_{D} = 
\begin{cases}
565 + 6.11 \times 10^{-2}T - \frac{1.14\times10^{7}}{T^{2}} & (T < 750 \,  \rm{^{\circ}}C)\\
1.0 & (T \geq 750 \, \rm{^{\circ}}C)
\end{cases}
\end{equation}

\noindent
The contribution to the volumetric swelling due to fission products is computed using the empirical Finlay model. The issue for modeling $\rm U_{3}Si_{2}$ is caused by its limited data availability. To overcome this issue, a cumulative burnup based model is provided by Finlay \cite{finlay_2004}. This model is based on the experimental results of swelling obtained from multiple irradiation experiments. Finlay’s data (fission density) was converted to FIMA using the heavy metal density of 10.735 $\rm g/cm^{3}$, which is 95\% of the theoretical heavy metal density \cite{finlay_2004, metzger2014model}. The volumetric strain due to fission products ($fps$) combines solid ($sfp$) and gaseous fission product ($gfp$) terms as shown in Eq. \ref{eq:vol_strain_u3si2}.
\begin{equation}
\label{eq:vol_strain_u3si2}
\begin{split}
        \left( \frac{\Delta V}{V_{0}} \right)_{fps} &= \left( \frac{\Delta V}{V_{0}} \right)_{sfp} +  \left( \frac{\Delta V}{V_{0}} \right)_{gfp} \\
        &=  (0.34392 \times Bu) + (3.8808 \times Bu^{2} + 0.45419 \times Bu)    \\
        &= 3.9909 \times Bu^{2} +  0.79811 \times Bu
\end{split}
\end{equation}
where the burnup is in units of FIMA.

%%%%%%%%%%%%% SiC/SiC
\subsection{SiC/SiC Models}
Chemical Vapor Infiltrated SiC/SiC Composite (CVI SiC/SiC) is employed as a fuel cladding material in this study. The following built-in BISON models for CVI SiC/SiC are utilized: thermal conductivity, specific heat, elasticity tensor, thermal expansion, and volumetric swelling.

In BISON, there are three thermal conductivity models for SiC/SiC: Koyanagi \cite{koyanagi2017}, Stone \cite{stone2015}, and Combined models.
\begin{comment} 
which one(s) are being used in the simulation(s)? KMP 
\end{comment}
The Koyanagi model is for unirradiated composite sic/sic and is derived from experimental thermal diffusivity data for tube-shaped samples. The Stone model is applicable to both irradiated and unirradiated composite sic/sic. The Stone model is unique in that it derives thermal conductivity as a function of temperature only, as opposed to the Koyanagi model, which gives thermal conductivity as a function of thermal diffusivity, specific heat, and density of composite sic/sic. The combined model allows one to incorporate the irradiation response from the Stone model into the Koyanagi model \cite{bison}. For the specific heat capacity (J/kg-K), the model of monolithic SiC is employed, and it is given by
\begin{equation}
\label{eq:c_p_monolithic}
    C_{P} = 925.65 + 0.3773T-7.9259 \times 10^{-5} T^{2} - \frac{3.1946\times 10^{7}}{T^{2}}
\end{equation}
where $T$ is the temperature (K). In this study, the thermal conductivity and specific heat capacity are user-defined values for the monolithic SiC model (Eq. \ref{eq:c_p_monolithic}). 

% Elasticity tensor
The Elasticity tensor of composite SiC/SiC is complex. In general, when SiC/SiC is used for cladding applications, it is considered to be orthotropic: 3D braided architecture with fibers along the axial direction and angles $ \pm \theta$ from the axial direction of the cladding \cite{bison}. The material properties are affected by fiber orientation and porosity. Experimental works are ongoing to quantify these effects. In this work, as suggested by literature \cite{sic}, we assumed Young's modulus of $90$ GPa and Poisson's ratio of $\nu = 0.35$.

% Thermal expansion
The strain due to thermal expansion is computed from the correlation reported by Koyanagi \cite{koyanagi2017}. The linear thermal expansion coefficient ($\rm K^{-1}$) is given as a function of temperature (K) as follows:

\begin{equation}
\alpha = 1.0 \times 10^{-6}(-0.7765 + 0.014350T - 1.2209 \times 10^{-5}T^{2} + 3.8289 \times 10^{-9}T^{3})
\end{equation}
where the valid temperature range is $294 < T \leq 1273$. The thermal strain is computed incrementally using the average coefficient of thermal expansion across the time step given the incremental temperature change \cite{bison}. 

\begin{equation}
    \epsilon_{th}^{curt} = \epsilon_{th}^{pre} + (T_{curt} - T_{pre}) \frac{\alpha_{curt} + \alpha_{pre}}{2}
\end{equation}
where the superscript $curt$ and subscript $pre$ represent the current time step and previous time step, respectively. The simulation's reference temperature for no thermal expansion event must be set in BISON, with a temperature of 295 (K) employed in this study.

%% Volumetric Swelling
BISON utilizes the same volumetric swelling model as the monolithic SiC for composite SiC/SiC. Katoh model is one option that includes both temperature and neutron fluence effects \cite{bison}. The swelling strain is expressed as a differential equation of fast neutron fluence for this option \cite{bison,stone2015}:

\begin{equation}
    \frac{dS}{d\gamma} = k(T) \gamma^{-1/3} \text{exp}\left( - \frac{\gamma}{\gamma_{sc}(T)} \right)
\end{equation}
where $S$ is the swelling strain, $\gamma$ is the fast neutron fluence ($\rm cm^{-2}$), $k(T)$ is a temperature-dependent rate constant, and $\gamma_{sc}(T)$ is a temperature-dependent characteristic fast neutron fluence. The following gives the two constants:

\begin{equation}
    k(T) = 6.0631 \times 10^{-8}T^{2} - 1.5904 \times 10^{-4}T + 0.10612
\end{equation}
and
\begin{equation}
    \gamma_{sc}(T) = 6.7221 \times 10^{-12}T^{4} - 1.3095 \times 10^{-8}T^{3} + 9.4807 \times 10^{-6}T^{2} - 2.7651 \times 10^{-3}T + 0.51801
\end{equation}

The Katoh model has an option for the number of steps in the incremental calculation of swelling in the low fluence region. In this study, it is set to 1000.

%%%
The summary of the built-in material models is listed in Table \ref{tb:model_all}.

%%%%%%%%%% Table
\begin{table}[htbp]
\centering
\caption{List of the BISON built-in material models that were used in this research}
\label{tb:model_all}
\begin{adjustbox}{max width=\textwidth}
\begin{threeparttable}

\begin{tabular}{@{}lllc@{}}
\toprule
Material                    & Physical quantity                        & BISON Function name                                     & Reference \\ \midrule
\multirow{6}{*}{$\rm UO_{2}$}    & Thermal conductivity \& heat capacity           & HeatConductionMaterial\tnote{1}                    & \cite{bison} \\
                         & Elasticity tensor         & UO2ElasticityTensor                       & \cite{bison} \\
                         & Stress                    & ComputeMultipleInelasticStress\tnote{2}   & \cite{bison} \\
                         & Creep                     & UO2CreepUpdate                            & \cite{hagrman93} \\
                         & Thermal expansion         & UO2ThermalExpansionMATPROEigenstrain      & \cite{hagrman93} \\
                         & Volumetric swelling       & UO2VolumetricSwellingEigenstrain          & \cite{hagrman93, rpt_rashid_2004} \\ \midrule
\multirow{6}{*}{$\rm U_{3}Si_{2}$} & Thermal conductivity \& heat capacity         & HeatConductionMaterial\tnote{1} &      \cite{bison}     \\
                         & Elasticity tensor         & U3Si2ElasticityTensor                     & \cite{u3si2_handbook2} \\
                         & Stress                    & ComputeMultipleInelasticStress\tnote{2}            & \cite{bison} \\
                         & Creep                     & U3Si2CreepUpdate                          &  \cite{yingling2019} \\
                         & Thermal expansion         & U3Si2ThermalExpansionEigenstrain          &  \cite{u3si2_handbook} \\
                         & Volumetric swelling       & U3Si2VolumetricSwellingEigenstrain        &  \cite{rpt_rashid_2004, finlay_2004, hofman1989} \\ \midrule
\multirow{5}{*}{SiC/SiC} & Thermal conductivity \& heat capacity              & HeatConductionMaterial\tnote{1}                    & \cite{bison} \\
                         & Elasticity tensor              & ComputeIsotropicElasticityTensor\tnote{3}          & \cite{malvern1969introduction, slaughter2012linearized} \\
                         & Stress               & ComputeStrainIncrementBasedStress         & \cite{bison} \\
                         & Thermal expansion    & CompositeSiCThermalExpansionEigenstrain   & \cite{koyanagi2017} \\
                         & Volumetric Swelling & CompositeSiCVolumetricSwellingEigenstrain & \cite{mieloszyk2015} \\ \bottomrule
\end{tabular}

\begin{tablenotes}
\small
\item[1]{the setting of user-defined thermal conductivity in the BISON input file}
\item[2]{iteratively compute stress, internal parameters, and plastic strains}
\item[3]{compute a constant isotropic elasticity tensor}
\end{tablenotes}
\end{threeparttable}
\end{adjustbox}
\end{table}

\subsection{Problem Setup}
\begin{figure}[!htbp]
    \centering
    \includegraphics[width=13cm]{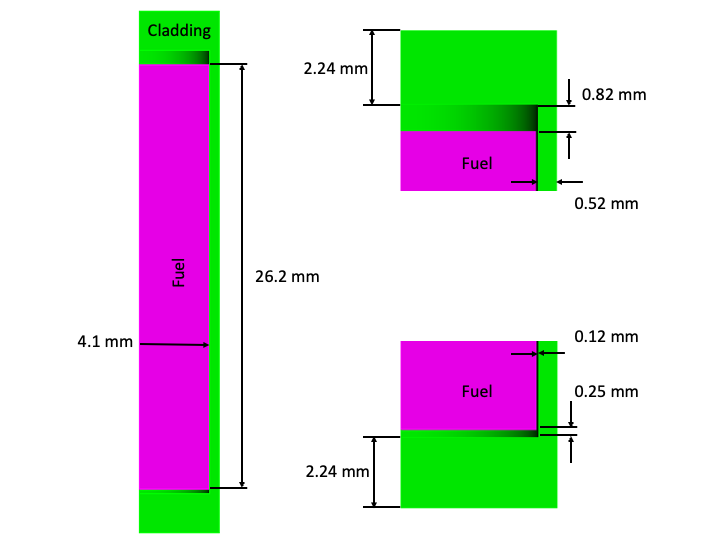}
    \caption{Illustration of a one-quarter fuel pin in 3D geometry, with the fuel pellet domain depicted in purple and the cladding highlighted in green.}
    \label{fig:fuel_rod_geo}
\end{figure}

 The model is composed of three regions: fuel, gap, and cladding. The system overview is shown in Figure \ref{fig:fuel_rod_geo}, and the geometric parameters are listed in Table \ref{tab:geometry}. The fuel pellet is $\rm UO_{2}$ or $\rm U_{3}Si_{2}$, the gap is helium gas, and the cladding material is SiC/SiC.
\begin{comment} 
Do you have the plenum region coupled with the fission gas production of UO2 to allow for the fission gasses to pollute the plenum? KMP 
\end{comment}
A heat transfer analysis of the system is performed. The parameters for the heat source and  coolant channel are listed in Table \ref{tab:input_params}. The simulation is performed using the Dirichlet boundary condition for the displacements. Two sets of pressure boundary conditions are applied: one from the coolant, which applies pressure to the exterior of the cladding, and another from the plenum pressure, which applies pressure to both the exterior and interior surfaces of the fuel and cladding interior \cite{inl_pressure_bc}.

\begin{comment} 
Are you using the coolant channel action? If so, what are the pitch and other parameters? KMP 
\end{comment}

%% fuel_rod_mesh

\begin{comment} 
Is this a FAST simulation, or ATF? Why not use typical dimensions and plennum/volume ratios? KMP 
\end{comment}

\begin{table}[!htbp]
\centering
\caption{Geometric parameters of fuel, gap, and cladding \cite{kobayashi2023uncertainty}}
\label{tab:geometry}
%\begin{adjustbox}{width=\textwidth}
\begin{tabular}{@{}llllll@{}}
\toprule
{Domain}      &  & {Parameters} &  &  & {Values (mm)} \\ \midrule
\multirow{2}{*}{(1) Fuel}     &  & radius              &  &  & 4.1                  \\
                          &  & height              &  &  & 26.2                 \\ \midrule
\multirow{3}{*}{(2) Gap}      &  & plenum              &  &  & 0.82                \\
                          &  & radial              &  &  & 0.12                \\
                          &  & axial               &  &  & 0.25                 \\ \midrule
\multirow{5}{*}{(3) Cladding} &  & inner radius        &  &  & 4.22                 \\
                          &  & outer radius        &  &  & 4.74                 \\
                          &  & radial thickness    &  &  & 0.52                \\
                          &  & axial thickness     &  &  & 2.24                 \\
                          &  & height              &  &  & 29.3                 \\ \bottomrule
\end{tabular}
%\end{adjustbox}
\end{table}

\begin{table}[htbp]
\centering
\caption{Input parameters for the problem: heat source and coolant}
\label{tab:input_params}
\begin{tabular}{@{}lll@{}}
\toprule
Parameters                & Values   & Units                         \\ \midrule
Fission energy            & $3.20 \times 10^{-11}$ & J/fission    \\
Coolant inlet temperature & 580      & K                           \\
Coolant inlet pressure    & 15.5     & MPa                         \\
Coolant mass flux         & 3800     & $\rm kg/m^{2} \mhyphen sec$ \\
Rod diameter              & $9.48 \times 10^{-3}$ & m                           \\
Fuel pin pitch                 & $1.26\times 10^{-2}$ & m                           \\ \bottomrule
\end{tabular}
\end{table}

\begin{figure}[!htbp]
    \centering
    \includegraphics[width=10cm]{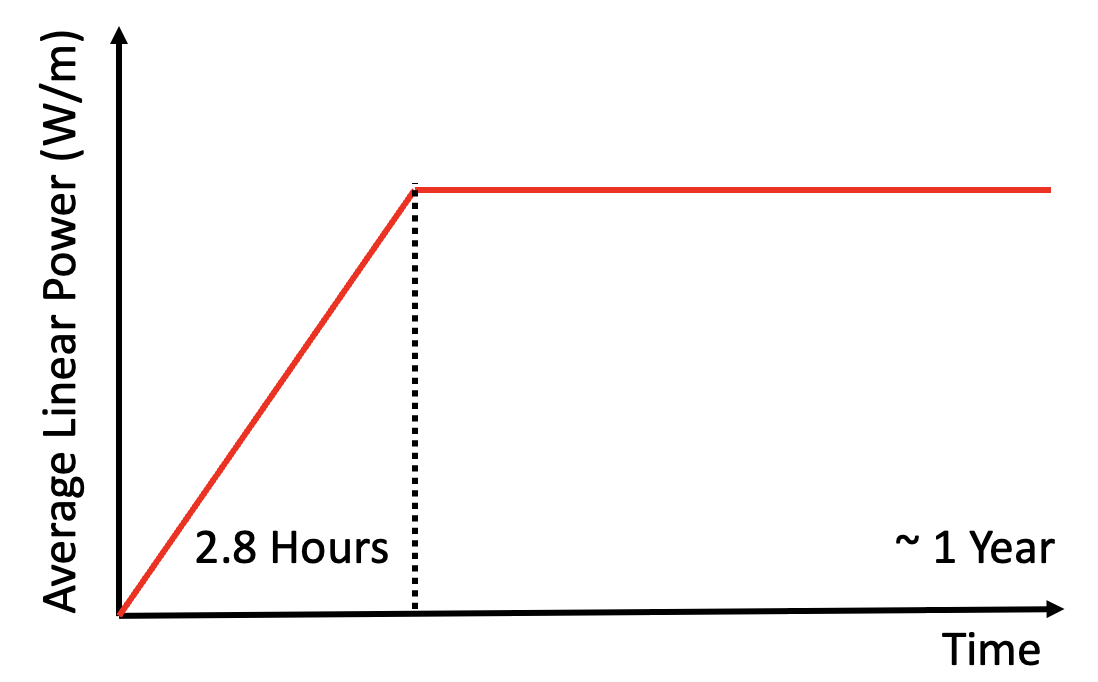}
    \caption{Power history for the simulation. The peak value is $1.5\times 10^{4}$ (W/m).}
    \label{fig:power_history}
\end{figure}

Figure \ref{fig:power_history} depicts the rod's power history under examination. It is characterized by an initial increase where the rod power ascends to its peak over 2.8 hours, followed by a sustained maximum for one year. This power profile has been selected to facilitate the demonstration of the non-intrusive UQ methods applied within the BISON framework. The selection of a 2.8-hour ramp-up period, along with the utilization of a simplified 3D geometry, is intended to streamline the complexity of the problem. 

% power history

%%%%%%%%%%%%%%%%%%%%%%%%%%%%%%%%

%%%%%
%%%%%

\subsection{Input Variables}
\label{sec:input}
In this example, the uncertain input parameters selected were the thermal conductivity and mass density of $\rm UO_{2}$ and SiC/SiC at room temperature. These parameters are critical as they influence outcomes related to temperature, burnup, and fission gas behaviors, necessitating their revision at every step of calculation. However, the primary objective of this investigation is to showcase the integration of UQ  methods within BISON. Therefore, to simplify the analysis, these input variables were designated as fixed values.

It was assumed that all input variables adhere to a Gaussian distribution characterized by a 5\% coefficient of variation, which is defined as the ratio of the standard deviation to the mean. The specific values for these parameters are detailed in Table \ref{tab:input_material}. It is important to address that the selection of higher thermal conductivity values for SiC/SiC composites is intentional and based on several forward-looking considerations pertinent to the rapid advancement/progress of ATFs for advanced reactors. This modeling approach is expected to be used in next-generation nuclear reactors, which will likely benefit from technological advances over the coming decades. We have assumed the highest quality manufacturing parameters available, taking into account anticipated advancements in manufacturing technology. The assumption of such high thermal conductivity is predicated on the utilization of: (a) High-quality SiC Fibers: Fibers with inherently high thermal conductivity like Sylramic or Tyranno SA. (b) High Matrix Density: Thorough CVI densification processes to minimize porosity and maximize the density of the SiC matrix, promoting higher thermal conductivity. (c) Optimized Fiber Architecture: Considered specific weave patterns or maximizing fiber volume fraction to enhance in-plane thermal conductivity. This aligns with the proactive/rapid development of ATFs, where performance characteristics are often pushed to their theoretical limits to understand potential gains in efficiency and safety.

\begin{table}[htbp]
\centering
\caption{List of input variables for fuels and SiC/SiC cladding \cite{kobayashi2023uncertainty}}
\label{tab:input_material}
\begin{adjustbox}{width=\textwidth}
\begin{tabular}{@{}llcc@{}}
\toprule
\multirow{2}{*}{Domain}   & \multirow{2}{*}{Input variables} & \multicolumn{2}{c}{Mean (RSD \%)}                        \\ \cmidrule(l){3-4} 
                          &                                  & \multicolumn{1}{c}{$\rm UO_{2}$} & \multicolumn{1}{c}{$\rm U_{3}Si_{2}$} \\ \midrule
\multirow{2}{*}{Fuel}     & Thermal conductivity (W/mK)  \cite{wu2022mechanism,antonio2018thermal}     & 2.8 (5)                           & 8.5 (5)                             \\
                          & Density ($\rm kg/m^{3}$)  \cite{wu2022mechanism,antonio2018thermal}                  & 10430.0 (5)                       & 11590.0 (5)                         \\
\multirow{2}{*}{SiC/SiC Cladding} & Thermal conductivity (W/mK)  \cite{kowbel2000high}    & \multicolumn{2}{c}{75 (5)}                                              \\
                          & Density ($\rm kg/m^{3}$)  \cite{kowbel2000high}                  & \multicolumn{2}{c}{2650.0 (5)}                                          \\ \bottomrule
\end{tabular}
\end{adjustbox}
\end{table}

\begin{figure}[!htbp]
    \centering
    \includegraphics[width=14cm]{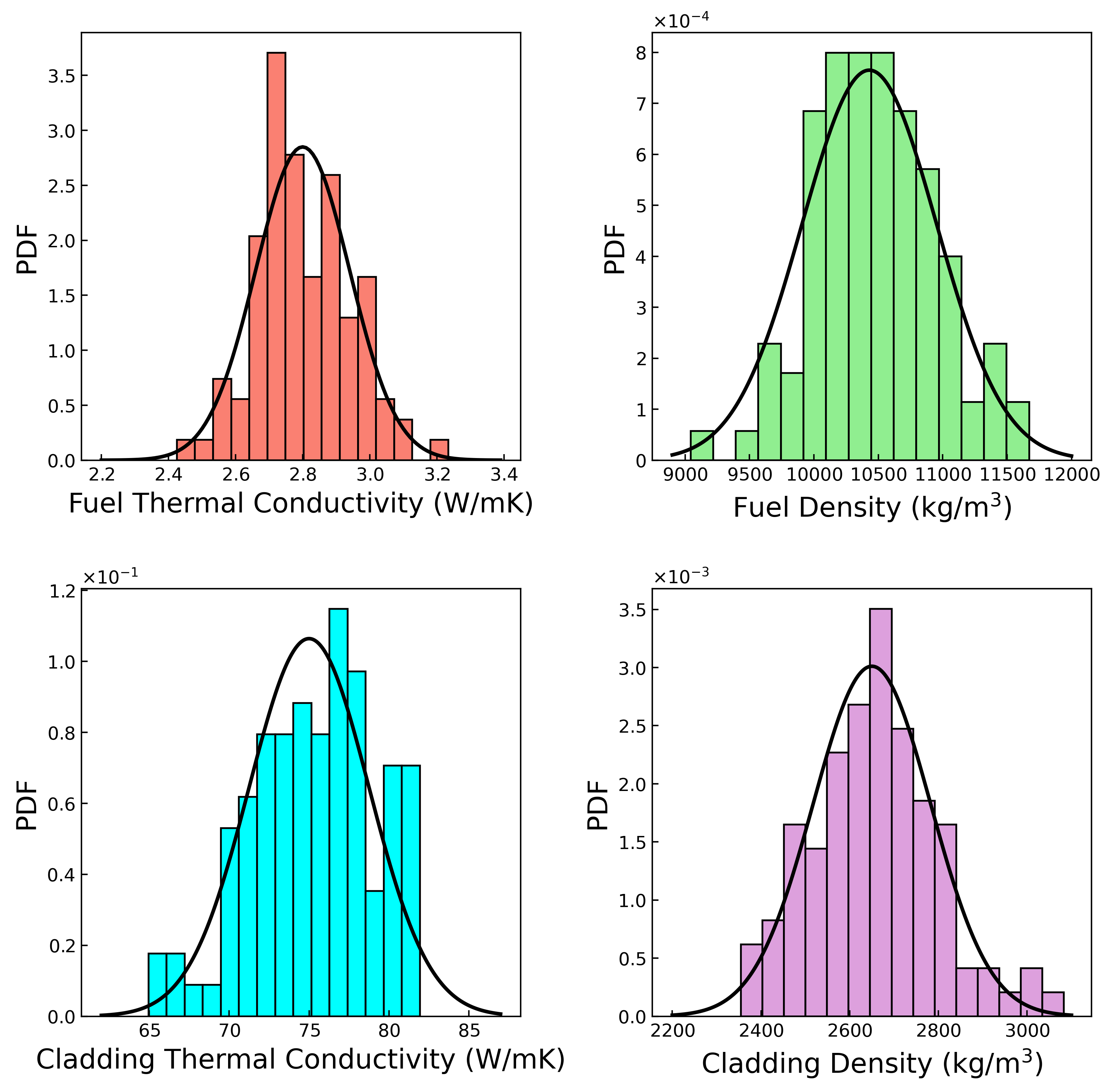}
    \caption{Probability distributions of input variables generated with the Monte Carlo sampling for $\rm UO_{2}+ SiC/SiC$ setup}
    \label{fig:uo2_input}
\end{figure}

As it is mentioned in Sections \ref{sec:mult_pols} and \ref{sec:regression}, the sampling number should be greater than $2(P+1)$ to ensure accuracy. In this problem, there are 4 input variables ($n$) and the polynomial order ($p$) of 3. Hence, the required sample number can be computed:

\begin{equation}
    2(P+1) = 2 \frac{(p+n)!}{p!n!} = 2 \frac{(3+4)!}{3!4!} = 70
\end{equation}
In order to ensure adequate over-sampling, 100 samples were generated for the $\rm UO_{2}$ system using the Monte Carlo sampling method. The PDE for $\rm UO_{2} + SiC/SiC$ is shown in Figure \ref{fig:uo2_input}.

% Input variable: UO2

% Input variable: U3si2
\begin{figure}[!htbp]
    \centering
    \includegraphics[width=14cm]{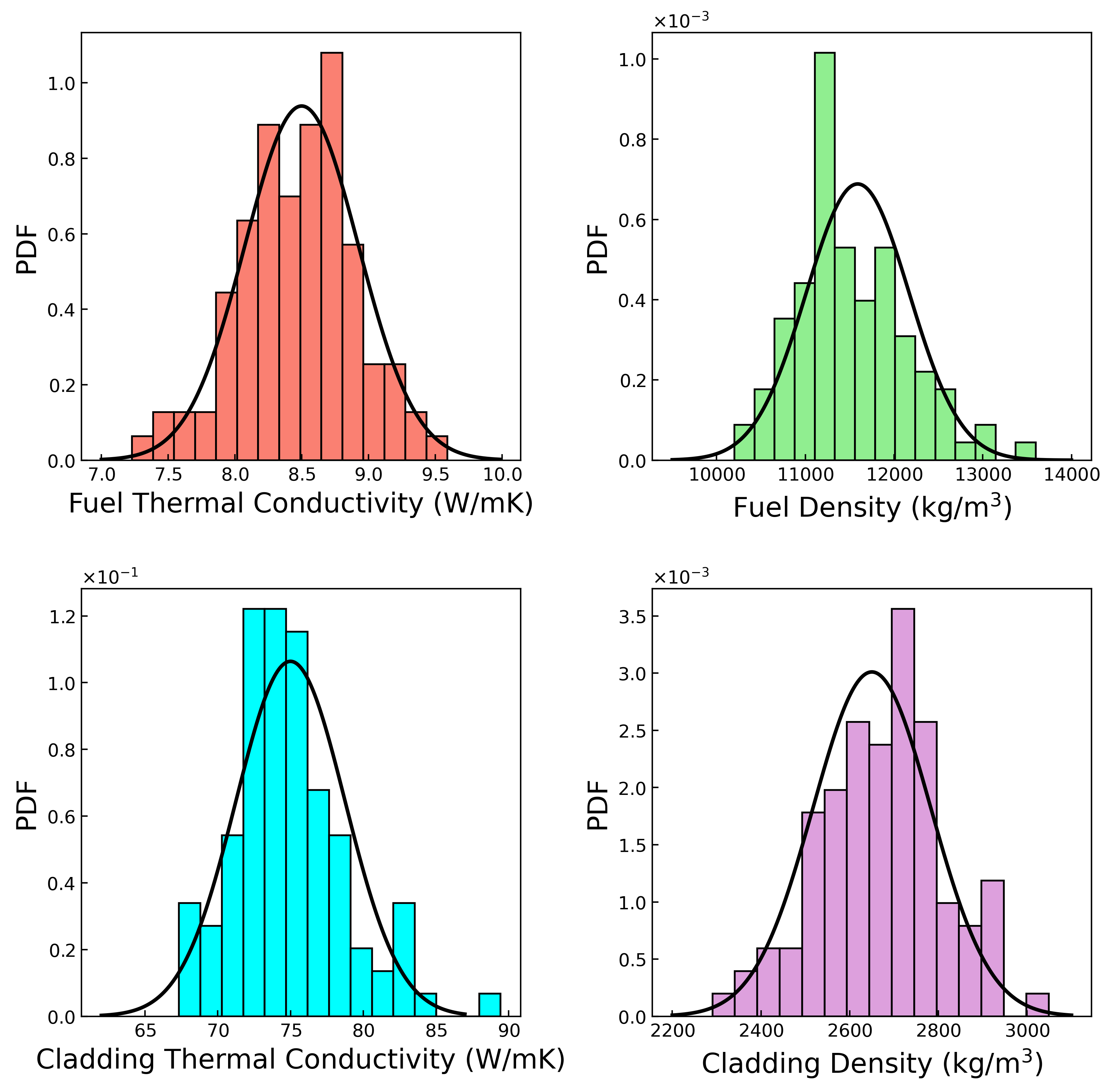}
    \caption{Probability distributions of input variables generated with the Monte Carlo sampling for $\rm U_{3}Si_{2}+ SiC/SiC$ setup}
    \label{fig:u3si2_input}
\end{figure}

\begin{figure}[!htbp]
    \centering
    \includegraphics[scale=0.6]{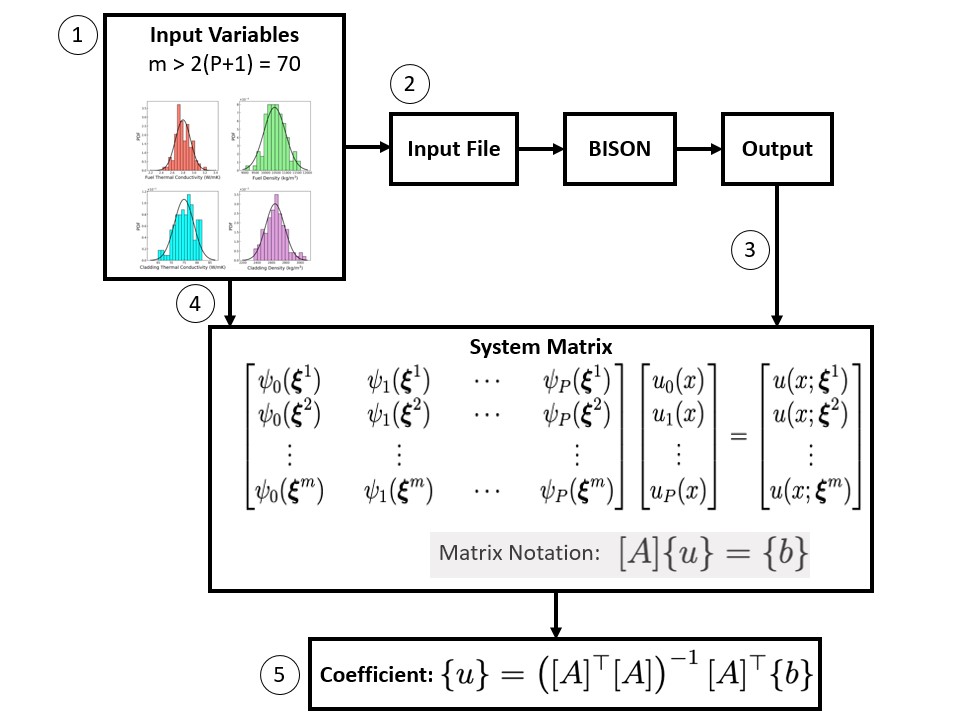}
    \caption{Flow chart of UQ analysis. }
    \label{fig:uq_matrix}
\end{figure}

% flow chart
Figure \ref{fig:uq_matrix} illustrates a series of tasks in the PCE methodology. The first task is the preparation of input variables, already described in Section \ref{sec:input}. The second task is to prepare input files for BISON. Since 100 sets of four input variable pairs are used in this study, a similar number of corresponding BISON input files are prepared. The third task is to feed the prepared input files to BISON and compile calculation results. The fourth task is to compute a multi-dimensional polynomial function from the random variables used in the input variable preparation (Section \ref{sec:mult_pols}). Tasks 3 and 4 yield the system matrix, and solving it for polynomial coefficients is the final task 5. The mean and variance of the output values are calculated using the coefficients and polynomial functions by Eqs. \ref{eq:mean_rephrased} and \ref{eq:var_rephrased}.

% power history

\section{Results and Discussion}
\begin{figure}[!htbp]
    \centering
    \includegraphics[width=15cm]{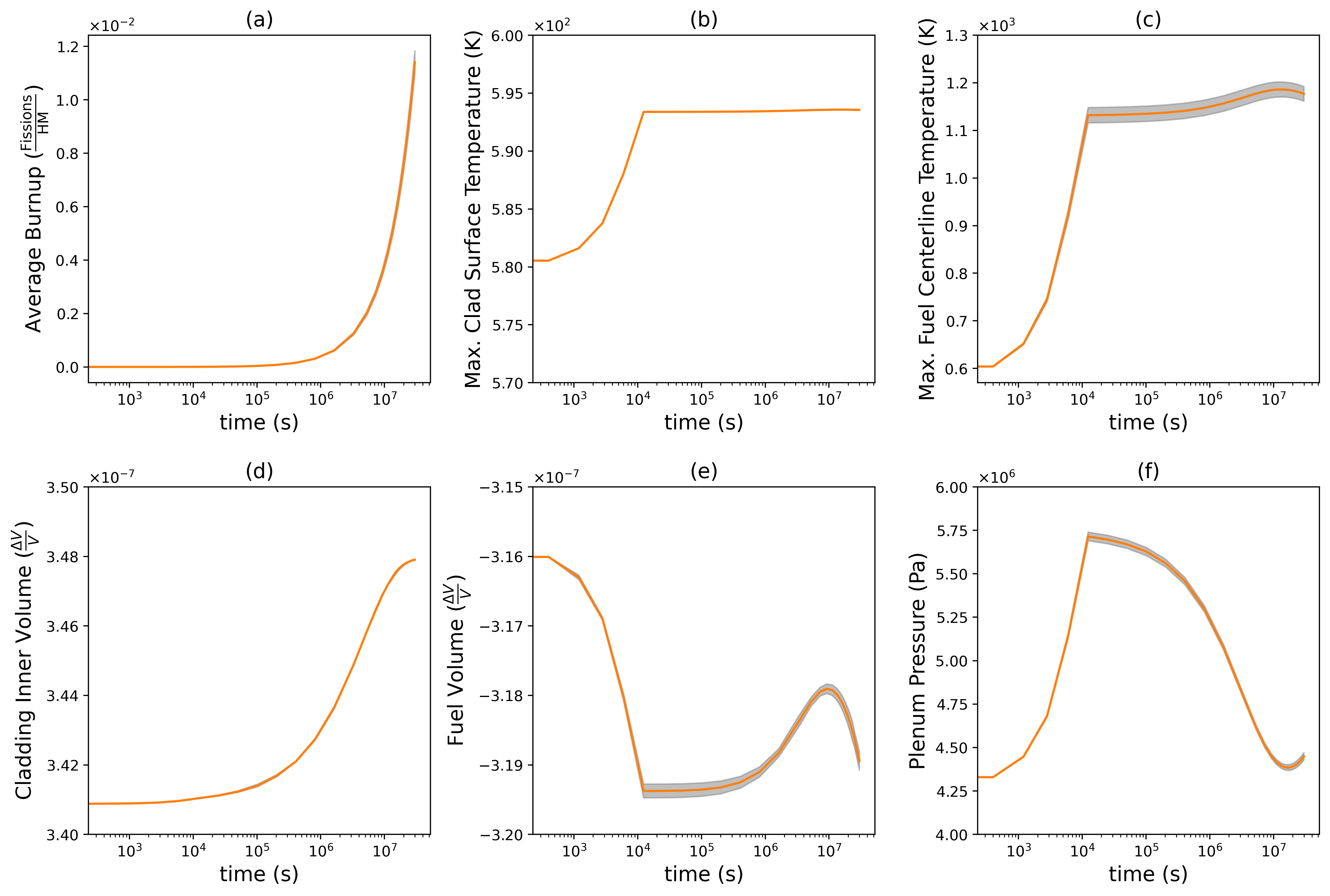}
    \caption{Results of the $\rm UO_{2}$+SiC/SiC. The mean values are shown with solid orange lines, and the confidence intervals corresponding to 1 standard deviation are drawn as shadow regions. For all output variables, the mean and standard deviation were computed.}
    \label{fig:uq_uo2}
\end{figure}

\begin{figure}[!htbp]
    \centering
    \includegraphics[width=15cm]{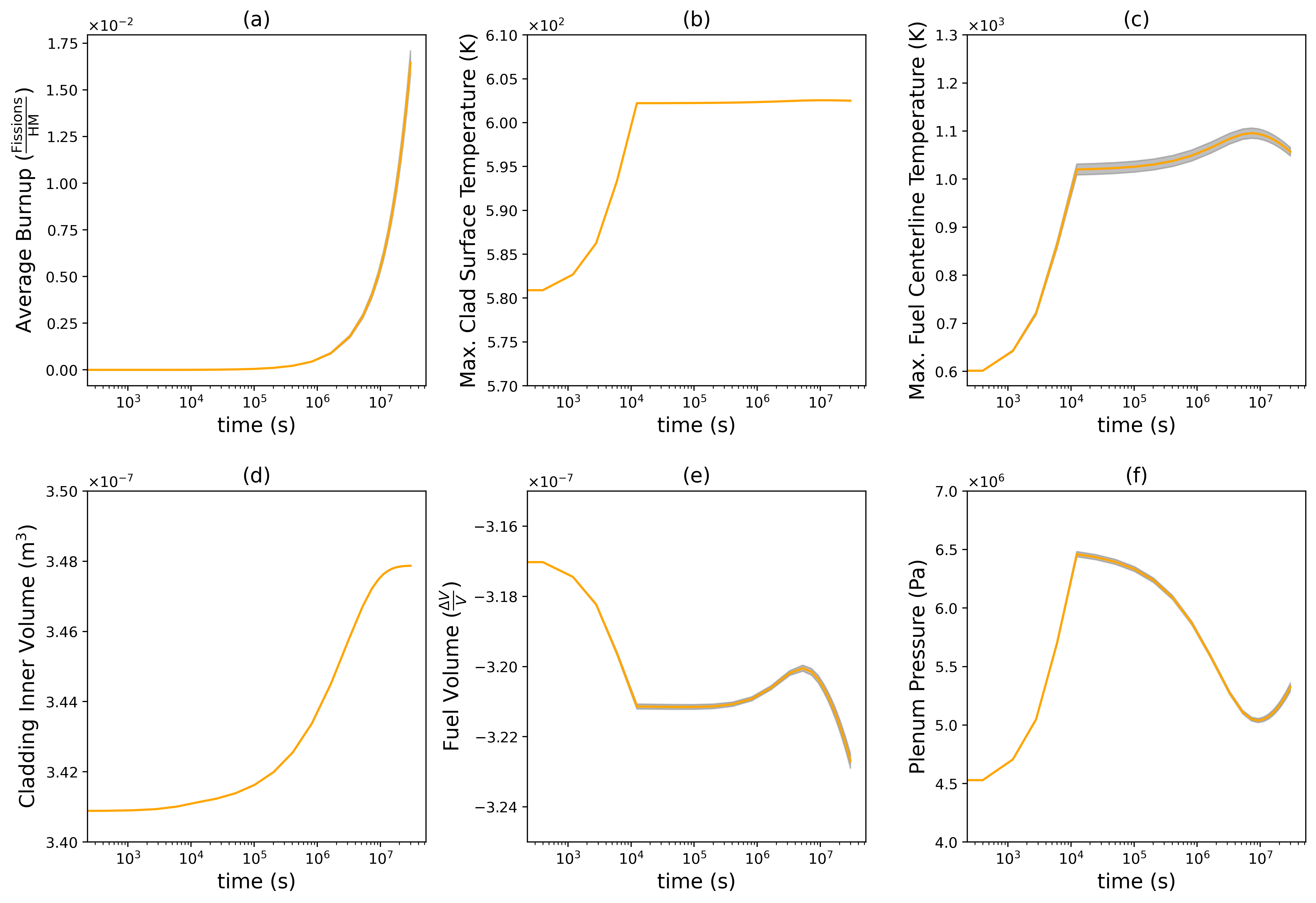}
    \caption{Results of the $\rm U_{3}Si_{2}$+SiC/SiC. The mean values are shown with solid orange lines, and the confidence intervals corresponding to 1 standard deviation are drawn as shadow regions. For all output variables, the mean and standard deviation were computed.}
    \label{fig:uq_u3si2}
\end{figure}

3D axisymmetric BISON simulations were performed to compute the average burnup, maximum cladding surface temperature, maximum fuel centerline temperature, cladding inner, fuel pellet volume, and plenum pressure. Figures \ref{fig:uq_uo2} and \ref{fig:uq_u3si2} represent the mean values and single standard deviation confidence intervals obtained with the PCE for $\rm UO_{2}$ and $\rm U_{3}Si_{2}$, respectively. 

% burnup
Figure \ref{fig:uq_uo2} (a) shows that the burnup and its uncertainty of the $\rm UO_{2}$ system increase slowly up to $10^{6}$ seconds, after which it increases rapidly with time additional.  Although the mean value differs, similar phenomena were observed in the $\rm U_{3}Si_{2}$ system. Figure \ref{fig:uq_u3si2} (a) represents 1.4 times higher mean values than $\rm UO_{2}$. This result is attributed to the high uranium density of $\rm U_{3}Si_{2}$ (11.3 $\rm g/cm^{3}$) \cite{white2015}. 

% clad surf & fuel cent temp
The cladding surface and fuel centerline temperatures of $\rm UO_{2}$ system are shown in Figure \ref{fig:uq_uo2} (b) and Figure \ref{fig:uq_uo2} (c). They appear to follow the power history. Both temperatures increase and reach equilibrium conditions at 2.8 hours. The cladding surface temperature's mean value and standard deviation converged ($\rm RSD < 0.0\%$). In contrast, the fuel centerline shows significant fluctuations ($\rm RSD \sim 1.2\%$). Also, it is confirmed that the maximum cladding surface and fuel centerline temperatures do not exceed their melting point, $\rm UO_{2}: 3138\,(K)$ and $\rm SiC: 3003\,(K)$. It remains true when the input uncertainties are taken into account. For the $\rm U_{3}Si_{2}$ system, the results are shown in Figure \ref{fig:uq_u3si2} (b) and Figure \ref{fig:uq_u3si2} (c). The $\rm U_{3}Si_{2}$ and $\rm UO_{2}$ systems show similar trends to each other as well as the burnup. It is noteworthy that the $\rm U_{3}Si_{2}$ system showed increased cladding temperature and decreased nuclear fuel temperature compared to the $\rm UO_{2}$ system. It is a typical example of the ATF concept of reducing fuel temperature by employing fuels with relatively high thermal conductivity. 

The volumetric changes in the cladding's inner region and the fuel pellet, depicted in Figures \ref{fig:uq_uo2} (d) and \ref{fig:uq_uo2} (e), exhibit a correlation that reflects the competitive interplay between densification and swelling due to solid and gaseous fission products. Initially, densification predominates, corresponding with the ramp-up in power. As operational conditions stabilize, a balance is achieved, leading to a plateau in volume change. Beyond $10^{7}$ seconds, a reversion to densification is observed, likely due to reduced fission gas generation as burnup stabilizes.

The inner volume of the cladding, while not significantly impacted by the uncertainty in input variables, shows sensitivity to cladding thermal conductivity, particularly influencing the cladding's response to heat flux and temperature gradients. For the fuel volume, thermal conductivity and density are primary influencers, affecting the fuel's thermal expansion and fission product retention.

Regarding plenum pressure, as illustrated in Figures \ref{fig:uq_uo2} (f) and \ref{fig:uq_u3si2} (f), a negative correlation with fuel volume suggests that as fuel densifies and volume decreases, the displaced gases increase plenum pressure. This effect is especially pronounced at critical junctures correlating with power history and fuel burnup milestones. The $\rm U_{3}Si_{2}$ system demonstrates a steeper pressure rise compared to $\rm UO_{2}$, indicative of its material properties and fission gas release behavior.Also, it is confirmed that plenum pressure is predominantly influenced by fuel density and thermal conductivity, underscoring the importance of these parameters in the design and operation of nuclear fuels.

\begin{comment}

\begin{figure}[!htbp]
    \centering
    \includegraphics[width=15cm]{SA_UO2.png}
    \caption{First order Sobol's sensitivity indices in the $\rm UO_{2}$ + SiC/SiC setup. The x-axis represents the index of input parameters, 1: fuel density, 2: fuel thermal conductivity, 3: cladding density, and 4: cladding thermal conductivity.}
    \label{fig:sa_uo2}
\end{figure}

\begin{figure}[!htbp]
    \centering
    \includegraphics[width=15cm]{SA_U3Si2.png}
    \caption{First order Sobol's sensitivity indices in the $\rm U_{3}Si_{2}$ + SiC/SiC setup. The x-axis represents the index of input parameters, 1: fuel density, 2: fuel thermal conductivity, 3: cladding density, and 4: cladding thermal conductivity.}
    \label{fig:sa_u3si2}
\end{figure}
\end{comment}

%% discussion
Only uncertainties of the input variables were taken into account in this work. In other words, the confidence intervals (standard deviations) depicted in Figure \ref{fig:uq_uo2} were likely underestimated since the code's inherent uncertainty was not taken into consideration. Notably, one must carefully consider the applicability of BISON's material modeling. Most built-in material models are prepared using empirical correlations, and their uncertainties are not utilized in the simulations. The model's validity is guaranteed in the case of established nuclear materials such as $\rm UO_{2}$, for which experimental data have been accumulated sufficiently. However, in simulations using innovative nuclear fuels and cladding materials, the validity and uncertainty of the empirical models will have a significant impact on the calculation results. Future research needs to quantify the impact on the final system outputs, assuming uncertainty in the input variables and the model.
\begin{comment} 
The discussion needs to be strengthened. Also, why not vary the scalars within the emperical material models themselves? KMP 
\end{comment}

\section{Conclusion}
Digital Twins integrated with AI/ML models can help NRC and DOE to make risk-informed decision-making for the Accelerated Fuel Qualification process for the advanced fuels. Among other components, uncertainty quantification analysis is the important segment of Digital Twins-enabling technologies to ensure trustworthiness. In this paper, uncertainty quantification method using polynomial chaos expansion was performed on the nuclear fuel performance code. As uncertain input variables, material densities and thermal conductivities for $\rm UO_{2}$/$\rm U_{3}Si_{2}$ fuels and SiC/SiC cladding were employed. It was demonstrated that the system output could be expressed in terms of mean and standard deviation even when uncertainties in the input variables are considered. It can be concluded that the methods presented in this study can provide more reliable calculation results to ensure the potential of BISON as a prediction tool in a Digital Twins for nuclear systems. 

Future work will focus on implementing prediction algorithms for BISON to tackle the ATF challenges of data unavailability, lack of data, missing data, and data inconsistencies. In addition, Explainable AI (XAI)-Infused Trustworthy Digital Twins framework and development of update module by Solving the “Inverse Problem” for synchronizing the physical and Digital Twins leveraging BISON code will be performed.

% \section*{Acknowledgement}
% The computational part of this work was supported in part by the National Science Foundation (NSF) under Grant No. OAC-1919789 and the High Performance Computing Center at Idaho National Laboratory, which is supported by the Office of Nuclear Energy of the U.S. Department of Energy and the Nuclear Science User Facilities under Contract No. DE-AC07-05ID14517.

\section*{Declaration of Generative AI and AI-assisted technologies in the writing process}
During the preparation of this work the author(s) used ChatGPT in order to language editing and refinement. After using this tool/service, the author(s) reviewed and edited the content as needed and take(s) full responsibility for the content of the publication. [\href{https://www.elsevier.com/about/policies/publishing-ethics/the-use-of-ai-and-ai-assisted-writing-technologies-in-scientific-writing}{Elsevier Publishing Ethics}]

\newpage 
\appendix
\section{Effect of fuel densification/swelling on the fuel centerline temperature}
\label{append:1}
\begin{figure}[!htbp]
    \centering
    \includegraphics[width=12cm]{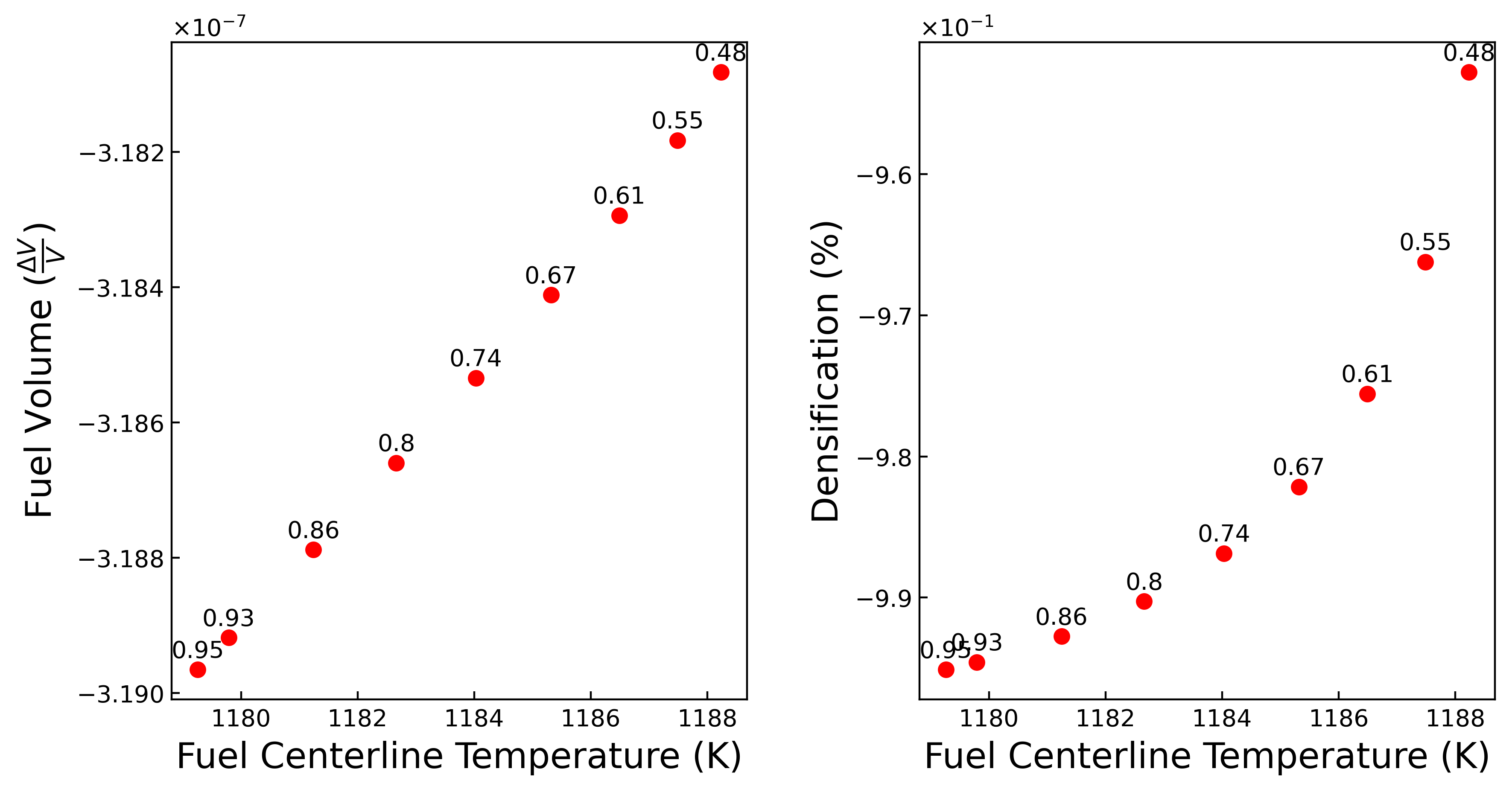}
    \caption{Correlations between the $\rm UO_2$ fuel centerline temperature and changes in fuel volume and densification. Annotations represent the operation time in years, highlighting the relationship over the period of the fuel's irradiation.}
    \label{fig:uo2_temp_density}
\end{figure}

Understanding the interplay between fuel densification/swelling and centerline temperature is crucial for accurately modeling nuclear fuel behavior during irradiation. Densification typically occurs during the initial phases of fuel life, often within the first year, and can significantly impact fuel performance.
This appendix analyzes the effects of fuel densification and swelling on the $\rm UO_2$/$\rm U_{3}Si_{2}$ fuel centerline temperatures. Figure \ref{fig:uo2_temp_density} illustrates the correlations between the $\rm UO_2$ fuel centerline temperature and changes in fuel volume and densification percentage. As indicated in the scatter plots of Figure \ref{fig:uo2_temp_density}, there is a noticeable trend where changes in fuel volume and densification percentage correlate with variations in fuel centerline temperature. The annotations on each data point represent the operation time in years, providing insight into the temporal aspect of these changes.

\begin{figure}[!htbp]
    \centering
    \includegraphics[width=12cm]{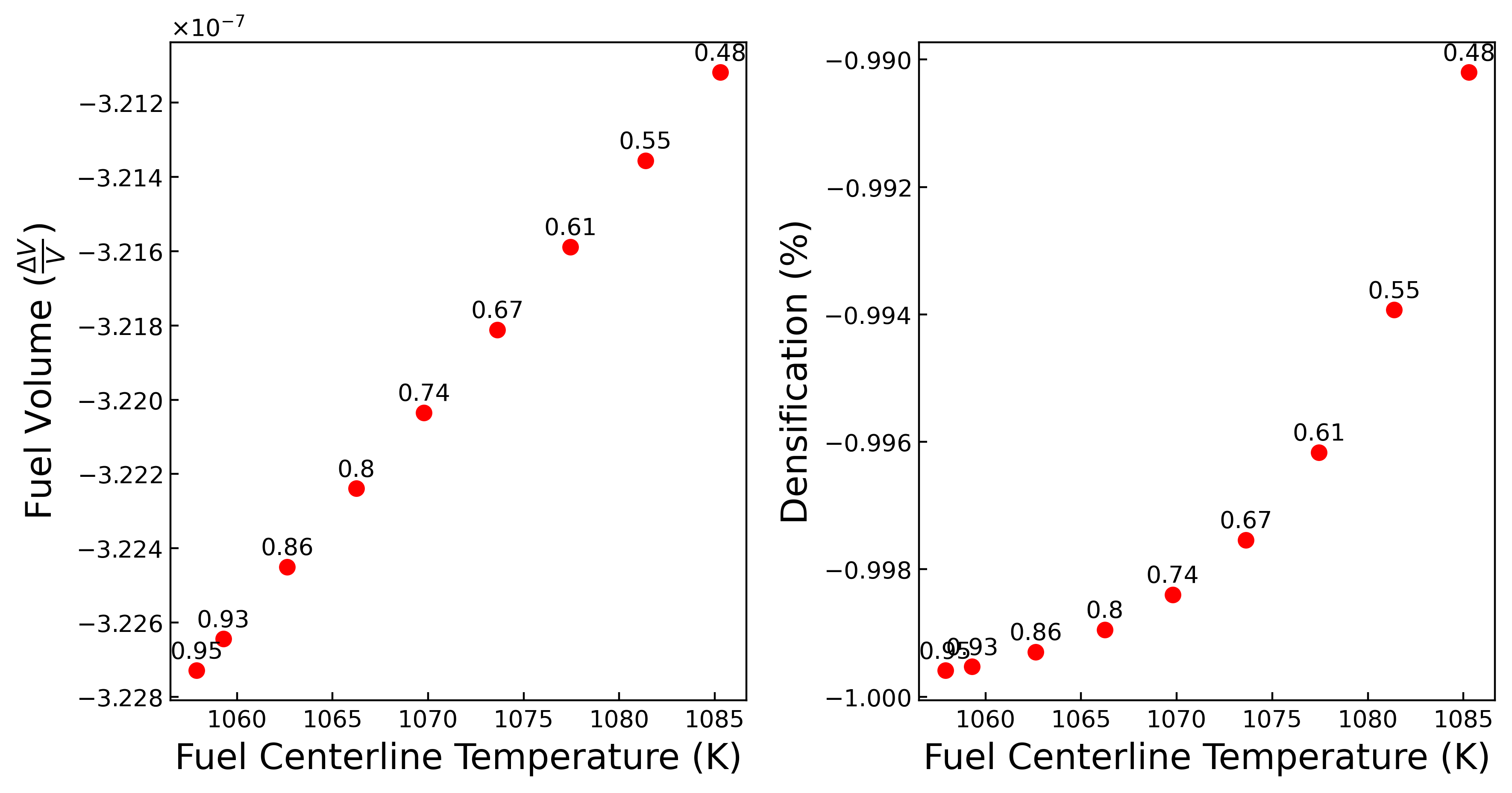}
    \caption{Correlations between the $\rm U_{3}Si_{2}$ fuel centerline temperature and changes in fuel volume and densification. Annotations represent the operation time in years, highlighting the relationship over the period of the fuel's irradiation.}
    \label{fig:u3si2_temp_density}
\end{figure}

Figure \ref{fig:u3si2_temp_density} presents an analogous study on  $\rm U_{3}Si_{2}$fuel. The scatter plot depicts a clear relationship between the centerline temperature of  $\rm U_{3}Si_{2}$ and its physical changes due to irradiation, notably densification and swelling.

As the figure shows, $\rm U_{3}Si_{2}$ demonstrates distinct behavior patterns in response to increasing centerline temperatures. Like $\rm UO_2$, the data points are annotated with the operational time in years, indicating a comprehensive timeline of the fuel's response to prolonged irradiation. This trend is crucial for predicting the long-term behavior of $\rm U_{3}Si_{2}$ in reactor conditions, particularly in scenarios aiming for extended fuel cycles or increased burn-up limits.

The comparative analysis between $\rm UO_2$ and  $\rm U_{3}Si_{2}$ fuels underscores the unique response of different fuel matrices to thermal and irradiation conditions, informing the optimization of fuel design and the advancement of nuclear fuel technology.

\bibliographystyle{unsrtnat}
\bibliography{references}  %%% Uncomment this line and comment out the ``thebibliography'' section below to use the external .bib file (using bibtex) .

\end{document}